Microstructure and transport properties of FeTe$_{0.5}$Se$_{0.5}$ superconducting wires fabricated by ex-situ Powder-in-tube process

T. Ozaki [a,b*], K. Deguchi [a,b], Y. Mizuguchi [a,b], H. Kumakura [a,b], Y. Takano [a,b]

[a] National Institute for Materials Science, 1-2-1 Sengen, Tsukuba, Ibaraki 305-0047, Japan

[a] JST, Transformative Research-project on Iron Pnictides, 1-2-1 Sengen, Tsukuba, Ibaraki 305-0047, Japan

**Abstract**

We fabricated FeTe$_{0.5}$Se$_{0.5}$ superconducting wires using ex-situ Powder-in-tube method with an Fe sheath. It was amazing that superconducting current was observed in the as-drawn wire without any heat treatments. By heat treatment at 200 °C for 2 hours, $T_c^{zero}$ and $J_c$ at 4.2 K were enhanced up to 9.1 K and 64.1 A/cm$^2$ ($I_c$ = 182.6 mA), respectively. Furthermore, the $J_c$ of FeTe$_{0.5}$Se$_{0.5}$ wire heat treated at 200 °C for 2 hours was not much sensitive to the applied magnetic fields. Therefore, FeTe$_{0.5}$Se$_{0.5}$ wires have a great potential for applications.

*(88 words)*

[*] Corresponding author.

Dr. Toshinori Ozaki

Postal address: National Institute for Materials Science, 1-2-1 Sengen, Tsukuba, Ibaraki 305-0047, Japan

Phone: +81-29-859-2644

Fax: +81-52-859-2601

E-mail address: OZAKI.Toshinori@nims.go.jp

## 1. Introduction

The discovery of iron-based superconductors [1] has caused great interest among the many researchers in superconductivity due to its high transition temperature $T_c$, even though they are mainly composed of iron, which has been considered to be detrimental to superconductivity because of its magnetism. Among the iron-based superconductors, The 11 series, such as FeSe, $FeTe_{1-x}Se_x$ and $FeTe_{1-x}S_x$, is an important iron-based superconducting system because the 11 series has the simplest crystal structures, and less toxicity compared to the other FeAs-based superconductors [2-4]. FeSe is significantly affected by applied pressure [4,5] and chalcogenide substitutions [6]. Furthermore, 11 series have a high upper critical field ($H_{c2}$) [4,7]. These results make 11 series promising candidates for technological applications.

We succeeded in the observation of a zero-resistivity current on the current-voltage measurement for the iron-based superconducting wires using the in-situ powder-in-tube (PIT) method with an Fe sheath [8]. However, obtained $J_c$ values in these wires are much lower than intra-grain $J_c$ [9]. Therefore, in order to enhance transport $J_c$, we attempted to fabricate $FeTe_{0.5}Se_{0.5}$ superconducting wires using the ex-situ PIT method. In this paper, we report on the effect of heat treatment on superconductivity and microstructure in the $FeTe_{0.5}Se_{0.5}$ wires.

## 2. Experimental

$FeTe_{0.5}Se_{0.5}$ wires were prepared using an ex-situ PIT method with an Fe sheath. $FeTe_{0.5}Se_{0.5}$ Polycrystalline samples prepared using the solid state reaction method were ground and tightly packed into a pure Fe tube with a length of 5 cm. This packing process was carried out in air. The inner and outer diameter of the tube were 3.5 and 6 mm, respectively. These tubes were cold-rolled into rectangular rods of 2.5 mm in size by using groove rolling and then drawn into a wire 1.1 mm in diameter. The round-shaped wire was cut into pieces of about 4 cm in length, and the wires were sealed into a quartz tube with an atmospheric-pressured argon gas. The sealed wires were rapidly heated at 150-500 °C for 2 hours.

Microstructure observations were performed using a scanning electron microscope (SEM). Temperature dependence of resistivity and transport critical current $I_c$ were measured by the four-probe method with a physical property measurement system (PPMS). The magnetic field was applied perpendicular to the wire axis. The criterion of $I_c$ definition was 1 $\mu$V/cm. The $J_c$ was obtained by dividing $I_c$ by the cross sectional area of the FeTe$_{0.5}$Se$_{0.5}$ core, which was measured by optical microscopy.

### 3. Result and Discussion

Fig. 1 shows the temperature dependence of resistivity at zero magnetic field for FeTe$_{0.5}$Se$_{0.5}$ wires fabricated by ex-situ PIT method. The values of $T_c^{onset}$, $T_c^{zero}$ and transition width ($\Delta T_c$) are listed in table 1. Surprisingly, superconducting current was observed in the as-drawn wire without any heat treatment. The zero-resistivity critical temperature ($T_c^{zero}$) was 3.2 K. In addition, with increase of heat treatment temperature, the values of $T_c^{onset}$ and $T_c^{zero}$ increased. The highest $T_c^{onset}$, $T_c^{zero}$ and $\Delta T_c$ were 10.8 K, 9.1 K and 1.7 K for the wire heat treated at 200 °C for 2 hours, respectively. The narrow $\Delta T_c$ is indicative of high homogeneity of the superconducting property. However, with an increase in temperature above 250 °C, the values of $T_c^{zero}$ were reduced. Finally, the FeTe$_{0.5}$Se$_{0.5}$ wire heat treated at 400 °C did not show zero-resistivity. We recently confirmed that Fe sheath reasonably supplied Fe for synthesizing the superconducting phase of Fe (Se,Te) [8]. Moreover, it is reported that the superconducting transition in single crystal of $Fe_yTe_xSe_{1-x}$ was suppressed by excess Fe [10]. Given these results, Fe sheath might excessively supply Fe for the superconducting phase, and the increase of excess Fe concentration resulted in reduced superconductivity. In addition, it is notable that the $T_c^{onset}$ values of each $FeTe_{0.5}Se_{0.5}$ wires were suppressed more than 4 K compared to that of $FeTe_{0.5}Se_{0.5}$ polycrystalline bulk [11]. This $T_c^{onset}$ reduction could be attributed to stress applied to the $FeTe_{0.5}Se_{0.5}$ phase during the cold-rolling process.

Fig. 2 shows Transport $J_c$ at 4.2 K and 2 K of $FeTe_{0.5}Se_{0.5}$ superconducting wires as a function of the heat treatment temperature. The typical current-voltage curve at self-field of the wire heat treated at 200 °C for 2 hours is shown in the inset. As shown in this inset, relatively sharp transition from superconducting to normal state is observed. The $J_c$ at 2 K of the as-drawn wire, which shows zero-resistivity at 3.2 K, exhibits a value of 2.8 A/cm$^2$. The $J_c$ increased with the increase of the heat treatment temperature until 200 °C. The transport $J_c$ values showed 82.5 A/cm$^2$ ($I_c$ = 235.0 mA) at 2 K and 64.1 A/cm$^2$ ($I_c$ = 182.6 mA) at 4.2 K, a factor of 5 higher than the value obtained Fe (Se,Te) tape fabricated by in-situ PIT method [8]. However, above 250 °C, the $J_c$ rapidly decreases with the increase of heat treatment temperature, and become zero above 300 °C. This is consistent with the tendency of the resistivity measurement shown in Fig.1.

In order to clarify the cause of the $T_c$ and $J_c$ improvement, the microstructure of $FeTe_{0.5}Se_{0.5}$ wires was investigated. Figs. 3(a) and 3(b) show the SEM images of the polished longitudinal cross section in $FeTe_{0.5}Se_{0.5}$ wires before heat treatment and after heat treatment at 200 °C for 2 hours, respectively. All the cross sections in Fig. 3 show uniform deformation of the composite without any problems such as breakage. However, there is an obvious difference in polished surface between before and after heat treatment. In order to reveal the detail of the composition, the higher magnification of cross section is shown in Fig. 4. It is evident that the FeTe$_{0.5}$Se$_{0.5}$ core presents a homogeneous cross section with no reaction layer between the superconducting core and the sheath. In FeTe$_{0.5}$Se$_{0.5}$ wire before heat treatment, although the microstructure of the FeTe$_{0.5}$Se$_{0.5}$ core was dense, the polished surface is very rough, indicating that the linkage of the grains cannot be so good. Therefore, obtained $T_c$ and $J_c$ values are very low. On the other hand, in FeTe$_{0.5}$Se$_{0.5}$ wire after heat treatment at 200°C for 2 hours, the polished surface of FeTe$_{0.5}$Se$_{0.5}$ cores is very smooth, suggesting better linkage of the FeTe$_{0.5}$Se$_{0.5}$ grain. It is considered that one of the reasons for the improved $T_c$ and $J_c$ of the wire heat treated at 200°C for 2 hours is the good linkage of the grains.

Magnetic field dependence of transport $J_c$ at 4.2 K for the FeTe$_{0.5}$Se$_{0.5}$ wire heat treated at 200°C for 2 hours was shown in Fig. 5. $J_c$-$B$ curve in Fe (Te,Se) tape using in-situ PIT method is also added for comparison [8]. An ex-situ FeTe$_{0.5}$Se$_{0.5}$ wire exhibited higher $J_c$ in all applied magnetic fields up to 7 T than an in-situ FeTe$_{0.5}$Se$_{0.5}$ tape. Moreover, decrease of the $J_c$ for ex-situ FeTe$_{0.5}$Se$_{0.5}$ superconducting wire with increasing magnetic field is low as compared to the one fabricated by in-situ PIT method. This result indicates that ex-situ PIT method would be effective for fabrication of FeTe$_{0.5}$Se$_{0.5}$ superconducting wire.

## 4. Conclusion

FeTe$_{0.5}$Se$_{0.5}$ superconducting wires were fabricated using the ex-situ PIT method with an Fe sheath. We observed superconducting current in the as-drawn wire without any heat treatments. In addition, by heat treatment at 200°C for 2 hours, $T_c^{zero}$ and $J_c$ were enhanced up to 9.1 K and 64.1 A/cm$^2$ ($I_c$ = 182.6 mA) at 4.2 K, respectively. This increase of the $J_c$ and $T_c$ by heat treatment is mostly due to the improved connection of FeTe$_{0.5}$Se$_{0.5}$ grains. Furthermore, the $J_c$ is not much sensitive to the applied magnetic fields. The optimization of the heat treatment conditions, together with the optimization of cold working of FeTe$_{0.5}$Se$_{0.5}$ wire, will lead to higher $J_c$ and $T_c$ values.

**Acknowledgments**

This work was supported by Japan Society for the Promotion of Science (22-10227).

**References**

[1] Y. Kamihara, T. Watanabe, M. Hirano, H. Hosono, J. Am. Chem. Soc. 130 (2008) 3296.

[2] F. C. Hsu, J. Y. Luo, K. W. Yeh, T. K. Chen, T. W. Huang, P. M. Wu, Y. C. Lee, Y. L. Huang, Y. Y. Chu, D. C. Yan, M. K. Wu, Proc. Natl. Acad. Sci. U.S.A. 105 (2008) 14262.

[3] M. H. Fang, H. M. Pham, B. Qian, T. J. Liu, E. K. Vehstedt, Y. Liu, L. Spinu, Z. Q. Mao, Phys. Rev. B 78 (2008) 224503.

[4] Y. Mizuguchi, F. Tomioka, S. Tsuda, T. Yamaguchi, Y. Takano, Appl. Phys. Lett. 93 (2008) 152505.

[5] S. Masaki, H. Kotegawa, Y. Hara, H. Tou, K. Murata, Y. Mizuguchi, Y. Takano, J. Phys. Soc. Jpn. 78 (2009) 063704.

[6] Y. Mizuguchi, F. Tomioka, S. Tsuda, T. Yamaguchi, Y. Takano, Appl. Phys. Lett. 94

(2009) 012503.

[7] T. Kida, M. Kotani, Y. Mizuguchi, Y. Takano, M. Hagiwara, J. Phys. Soc. Jpn. 78 (2009) 113701.

[8] Y. Mizuguchi, Keita Deguchi, S. Tsuda, T. Yamaguchi, H. Takeya, H. Kumakura, Y. Takano, Appl. Phys. Express 2 (2009) 083004.

[9] T. Taen, Y. Tsuchiya, Y. Nakajima, T. Tamegai, Phys. Rev. B 80 (2009) 092502.

[10] T. J. Liu, X. Ke, B. Qian, J. Hu, D. Fobes, E. K. Vehstedt, H. Pham, J. H. Yang, M. H. Fang, L. Spinu, P. Schiffer, Y. Liu, Z. Q. Mao, Phys. Rev. B 80 (2009) 174509.

[11] T. Ozaki, K. Deguchi, Y. Mizuguchi, H. Kumakura, Y. Takano, IEEE Trans. Appl. Supercond. (in press).

Table 1 $T_c^{onset}$, $T_c^{zero}$ and $\Delta T_c$ for FeTe$_{0.5}$Se$_{0.5}$ superconducting wires heat treated at different temperatures $T$ = 150-500 °C.

| Heat treatment temperature [°C] | $T_c^{onset}$ [K] | $T_c^{zero}$ [K] | $\Delta T_c$ [K] |
|---|---|---|---|
| as-drawn | 6.8 | 3.2 | 3.6 |
| 150 | 10.0 | 8.3 | 1.7 |
| 200 | 10.8 | 7.1 | 1.7 |
| 250 | 11.4 | 8.0 | 3.4 |
| 300 | 11.0 | 2.4 | 8.4 |
| 400 | 9.1 | — | — |
| 500 | 7.2 | — | — |

**Figure captions**

Fig. 1 Temperature dependence of resistivity for $FeTe_{0.5}Se_{0.5}$ superconducting wires heat treated at different temperatures $T = 150\text{-}500\,^\circ C$.

Fig. 2 $J_c$ values at 4.2 K and 2 K as a function of the heat treatment temperature for $FeTe_{0.5}Se_{0.5}$ superconducting wires. The inset shows the typical current-voltage curve at self-field of the $FeTe_{0.5}Se_{0.5}$ wire heat treated at $200\,^\circ C$ for 2 hours.

Fig. 3 Scanning electron micrographs on the longitudinal cross section of the $FeTe_{0.5}Se_{0.5}$ superconducting wires (a) before heat treatment and (b) after heat treatment at $200\,^\circ C$ for 2 hours.

Fig. 4 High-magnification Scanning electron micrographs on the longitudinal cross section of the $FeTe_{0.5}Se_{0.5}$ superconducting wires (a) before heat treatment and (b) after heat treatment at $200\,^\circ C$ for 2 hours.

Fig. 5 Magnetic field dependence of $J_c$ at 4.2 K for the $FeTe_{0.5}Se_{0.5}$ superconducting wire after heat treatment at $200\,^\circ C$ for 2 hours. $J_c$-$B$ curve in Fe (Te,Se) tape using in-situ PIT method is also added for comparison [8].

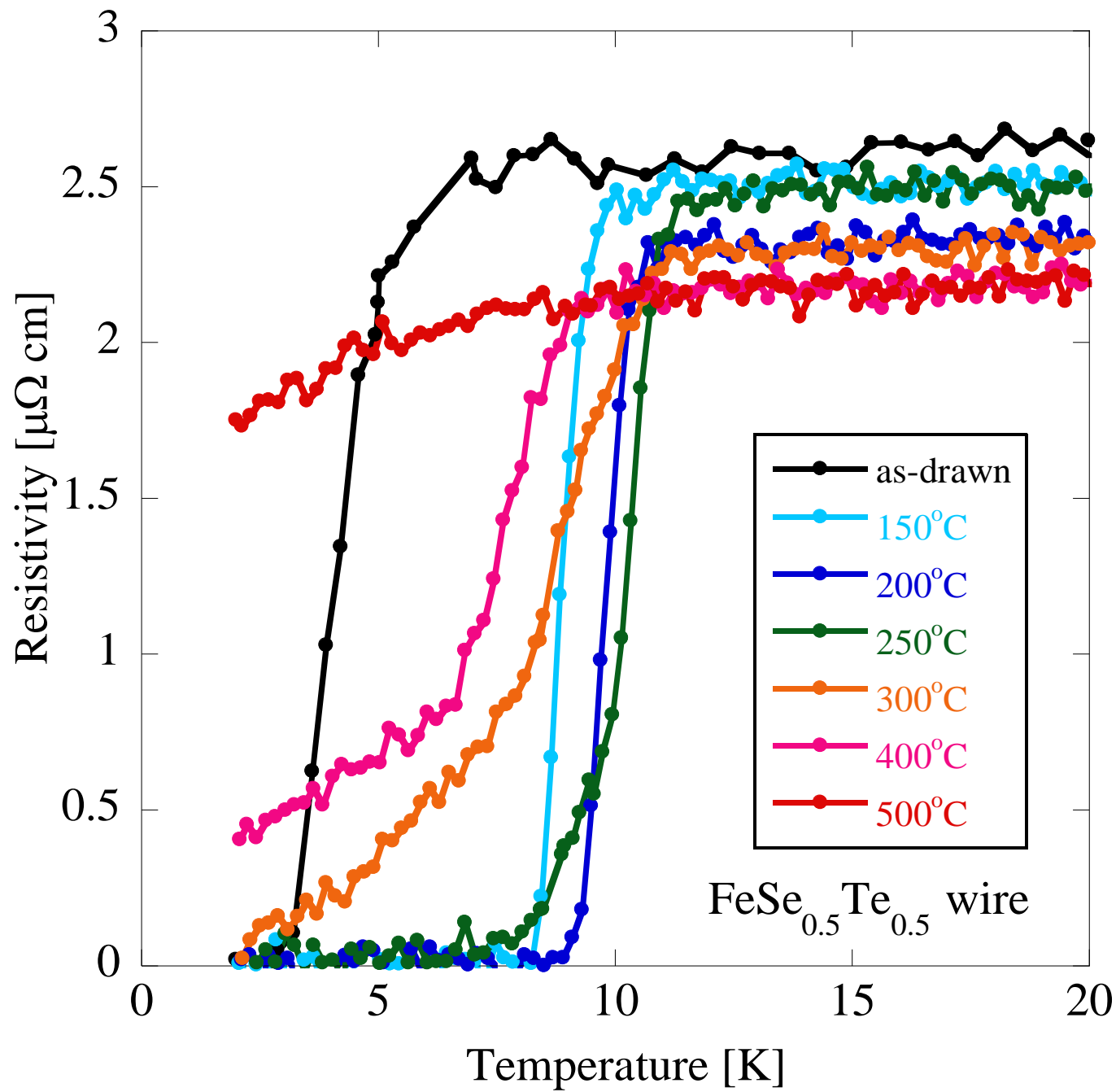

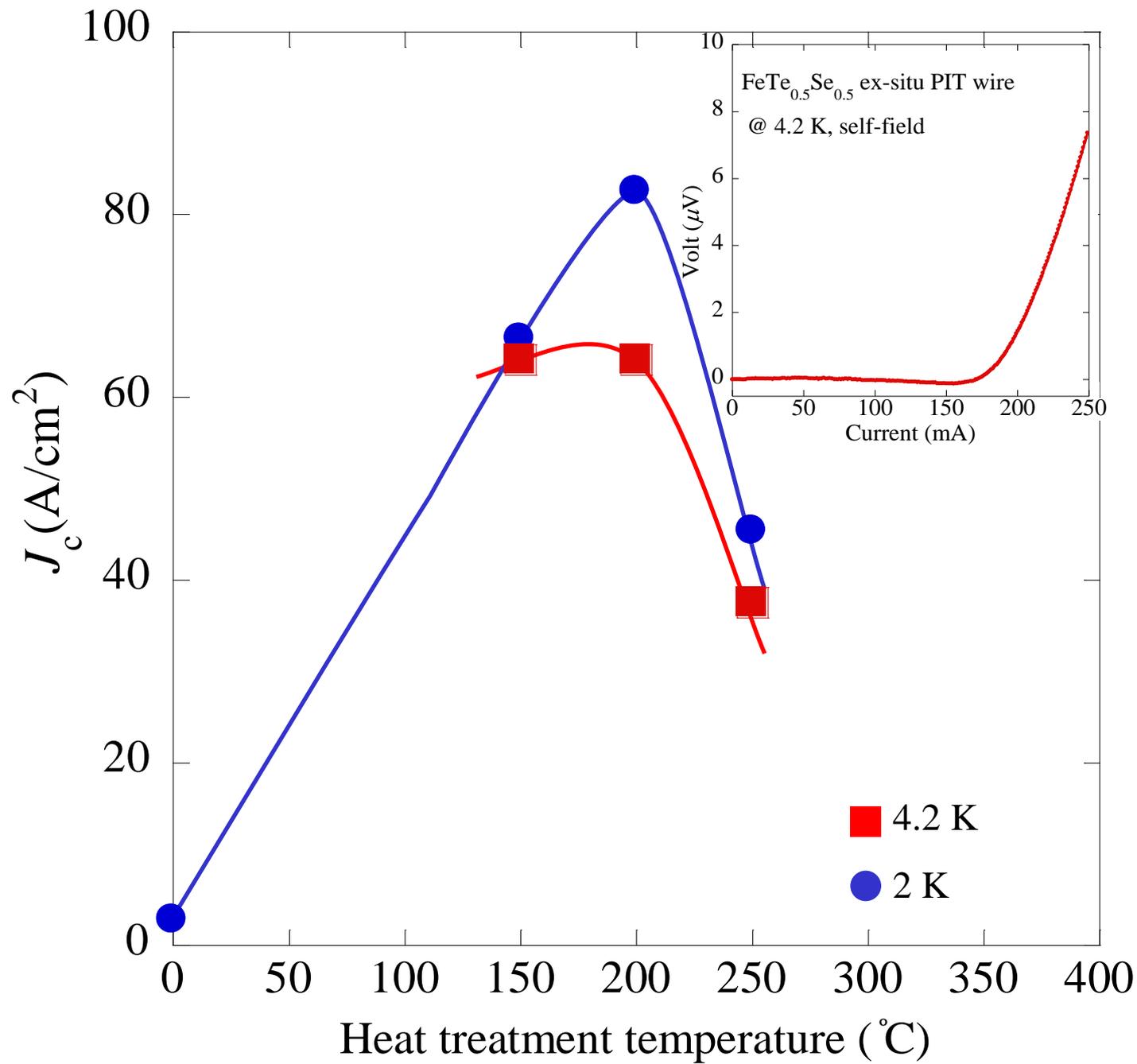

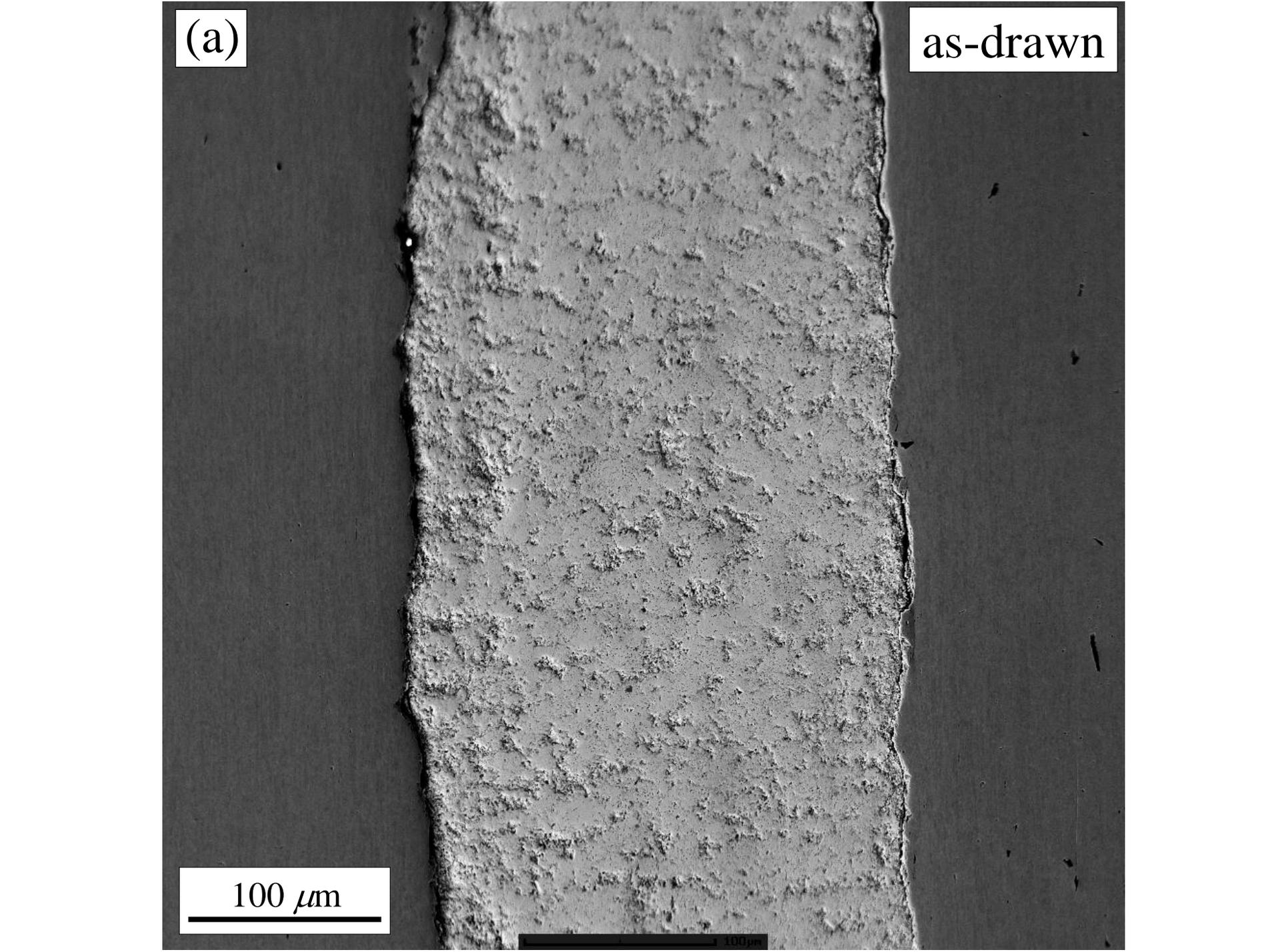

(a) as-drawn

100 μm

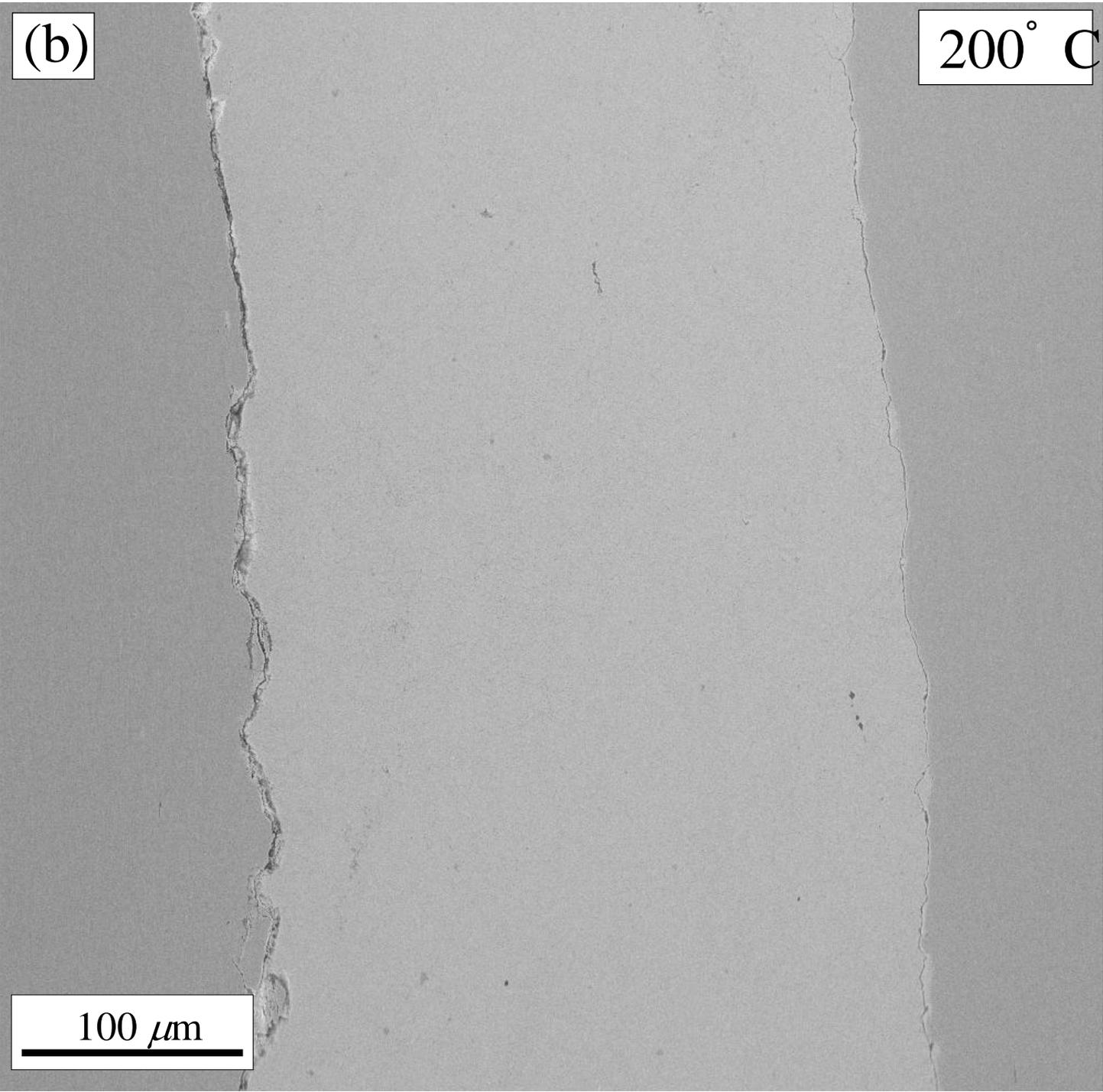

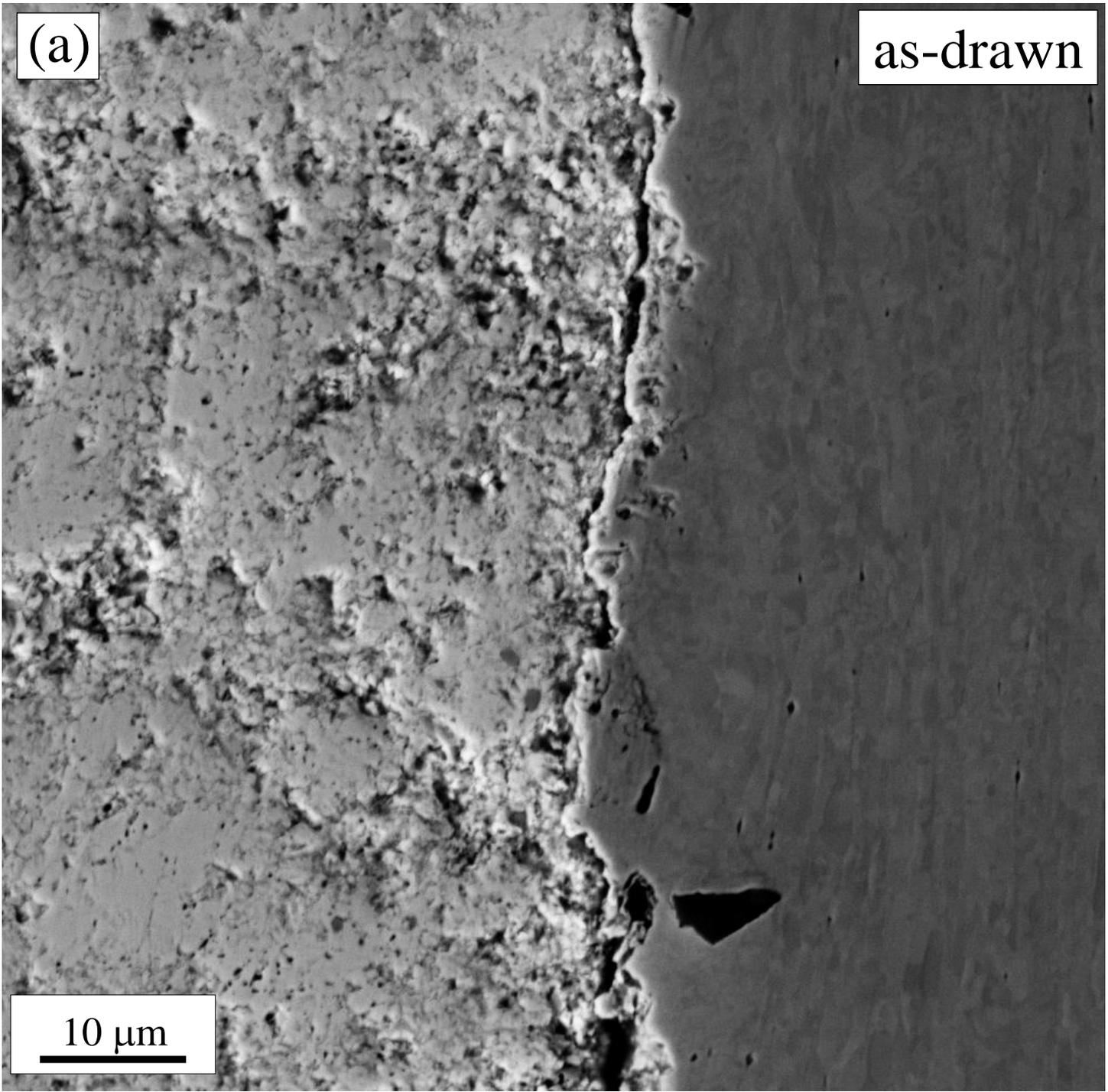

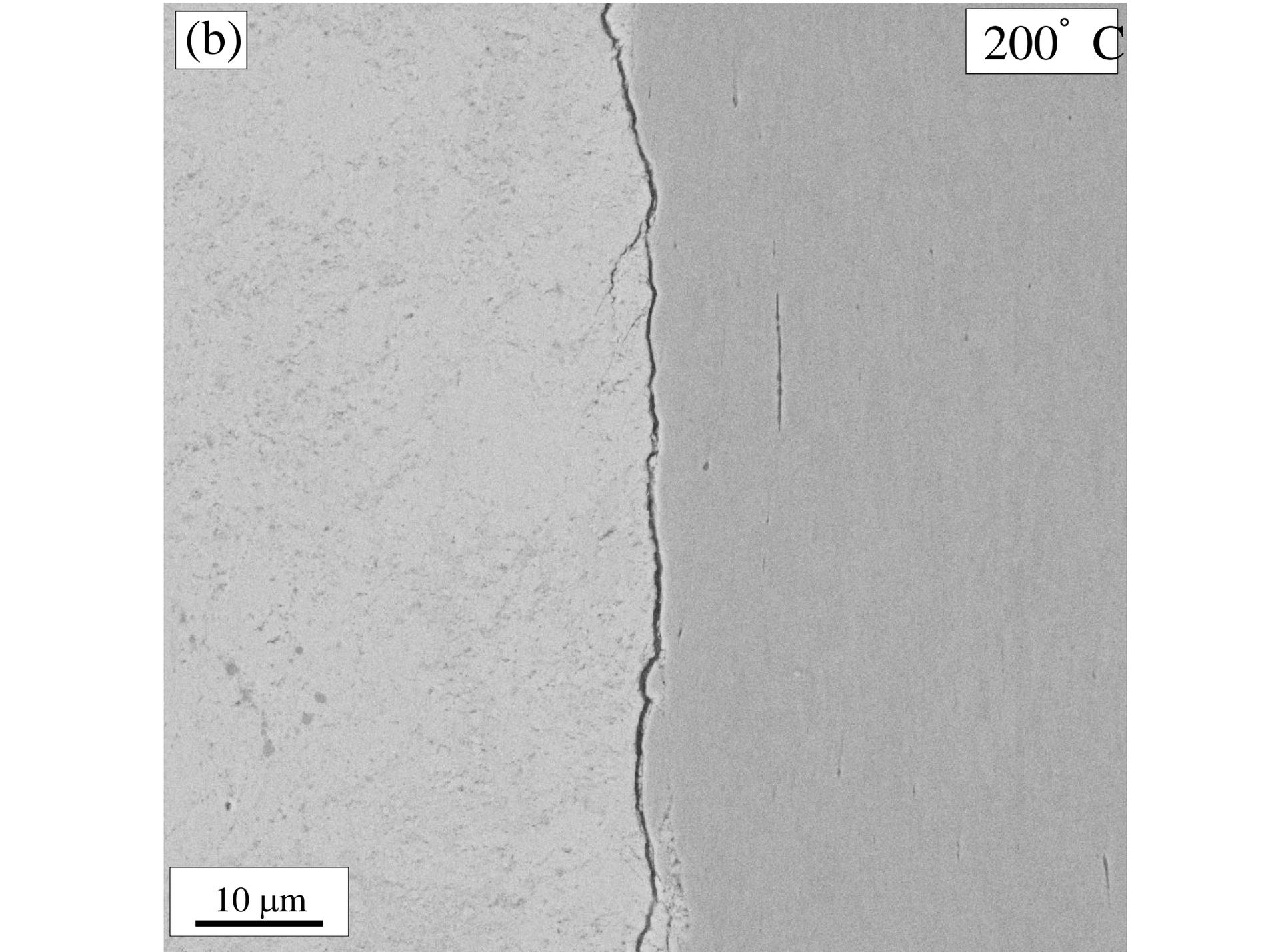

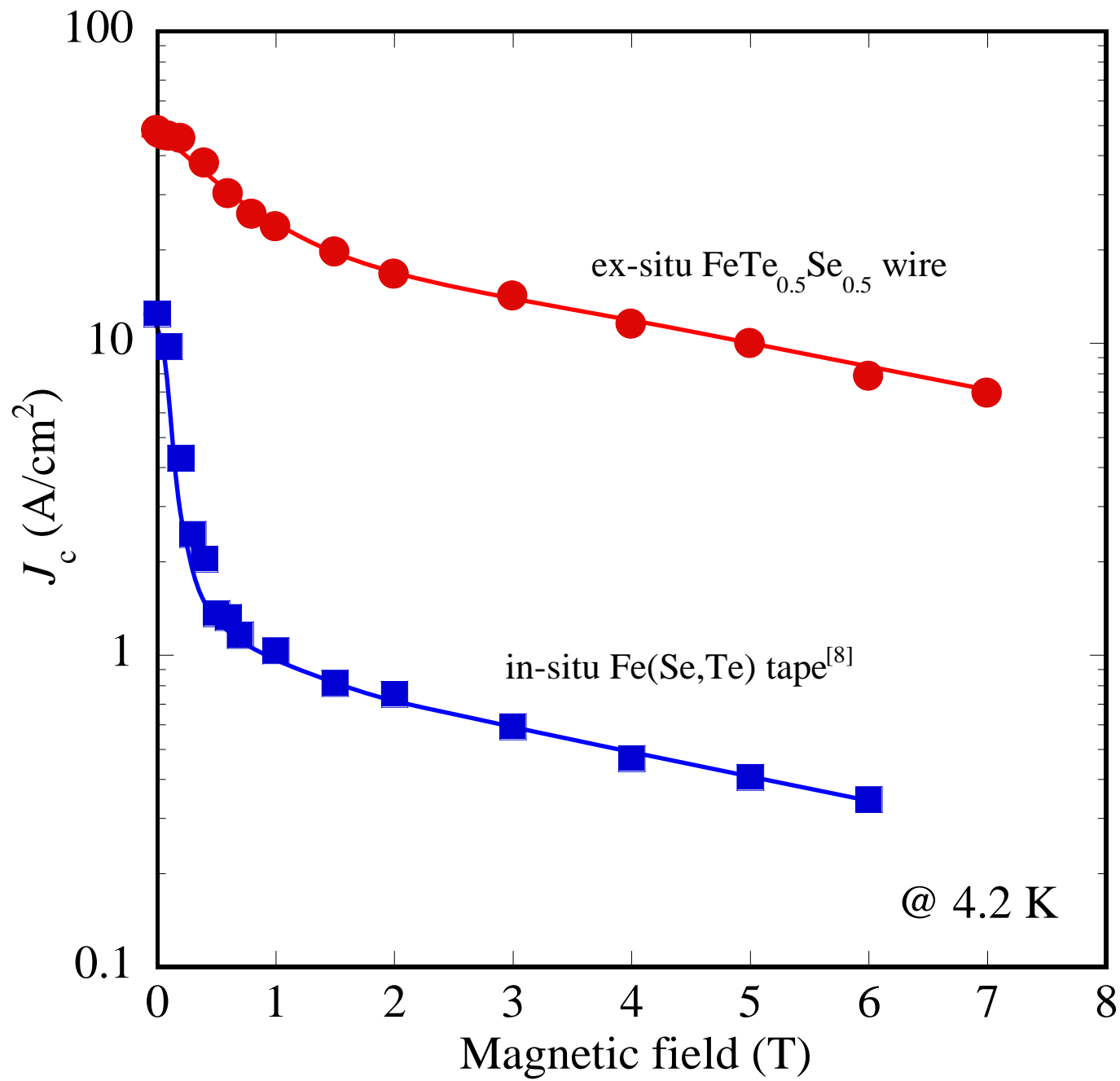